# Measurement independent magnetocaloric effect in Mn-rich Mn-Fe-Ni-Sn(Sb/In) Heusler alloys

Arup Ghosh[1, *], Rajeev Rawat[2], Arpan Bhattacharyya[3], Guruprasad Mandal[4], A. K. Nigam[5] and Sunil Nair[1,6]

[1]*Department of Physics, Indian Institute of Science Education and Research, Dr. Homi Bhabha Road, Pune 411008, India.*

[2]*UGC-DAE Consortium for Scientific Research, University Campus, Khandwa Road, Indore 452001, India.*

[3]*Saha Institute of Nuclear Physics, 1/AF Block Bidhannagar, Kolkata 700064, India.*

[4]*Centre for Rural & Cryogenic Technologies, Jadavpur University, Kolkata 700032 India.*

[5]*Department of Condensed Matter Physics and Materials Science, Tata Institute of Fundamental Research, Homi Bhabha Road, Mumbai 400005, India.*

[6]*Centre for Energy Science, Indian Institute of Science Education and Research, Dr. Homi Bhabha Road, Pune 411008, India.*

[*]Corresponding author's E-mail : arup@iiserpune.ac.in and arup.1729@gmail.com

**Abstract**

We report a systematic study on the magneto-structural transition in Mn-rich Fe-doped Mn-Fe-Ni-Sn(Sb/In) Heusler alloys by keeping the total valence electron concentration ($e/a$ ratio) fixed. The martensitic transition (MT) temperature is found to shift by following a proportional relationship with the $e/a$ ratio of the magnetic elements alone. The magnetic entropy change ($\Delta S_M$) across MT for a selected sample ($Mn_{49}FeNi_{40}Sn_9In$) has been estimated from three different measurement methods (isofield magnetization ($M$) vs temperature ($T$), isothermal $M$ vs field ($H$) and heat capacity ($HC$) vs $T$). We observed that though the peak value of $\Delta S_M$ changes with the measurement methods, the broadened shape of the $\Delta S_M - T$ curves and the corresponding cooling power (~140 Jkg$^{-1}$) remains invariant. The equivalent adiabatic temperature change ($\Delta T$) ~ -2.6 K has been obtained from indirect measurements of $\Delta T$. Moreover, an exchange bias field ~ 783 Oe at 5 K and a magnetoresistance of -30% are also obtained in one of these materials.

# 1. Introduction

Ni-Mn based full Heusler alloys are found to show large magnetocaloric effect (MCE) rear room temperature with high cooling power [1–3], which make them a potential candidate for environment friendly magnetic refrigeration technology. Heusler alloys can undergo a magneto-structural phase transition from ferromagnetic cubic austenite to weakly magnetic tetragonal, orthorhombic or monoclinic martensite structure. This first order magneto-structural transition (FOMST) involves significant changes in the thermal, magnetic, elastic and transport properties. Apart from the MCE, they are also known to show giant magnetoresistance (MR) [4–7], magneto-thermal conductivity [8], magnetostriction [9,10], etc. Most of these multifunctionalities are connected with the change in structural and magnetic properties across the FOMST. The martensite phase below 100 K is mostly found in mixed magnetic states and hence exhibit a large exchange bias (EB) effect in the bulk form [11–14].

Till date, a large number of Ni-Mn based Heusler alloys have been investigated for their MCE, MR and EB [2,3]. Addition of other transition metals like Co and Fe are found to be very effective in tuning the FOMST, enhancing the ferromagnetic correlations and consequently the MCE in them in bulk [7,14,23,24,15–22] and ribbon form as well [25,26]. Although, it is reported that the FOMST of these materials depends on the total valence electron concentration ($e/a$ ratio) of the composition and 3d states hybridization between Mn and Ni atoms, there are some exceptions where the conventional $e/a$ ratio dependence fails [1,3,27–32].

Apart from these, there are debates regarding the validity of Maxwell's equation for first order phase transitions. It is thought that one might observe a discontinuity in $dM/dT$ across the FOMST [2,3,33]. But in practical cases, the FOMST of Heusler alloys are not that sharp to show such discontinuity and thus the Maxwell's equation can still be used. However, the estimated values of MCE parameters like, isothermal magnetic entropy change ($\Delta S_M$) and refrigerant capacity (RC) necessarily need to be crosschecked from different measurement methods like, magnetization and heat capacity.

In an earlier work we have observed that the FOMST temperature follows the total *e/a* ratio in Fe doped Mn-rich Mn-Fe-Ni-Sn alloys where Fe was doped separately in the place of Ni and Mn [22]. In the present work, we have added Sb and In by replacing Sn in Fe doped Mn-rich Mn-Fe-Ni-Sn(Sb/In) alloys to keep the total *e/a* ratio fixed and then investigated the variation of the FOMST temperature, MCE, EB and MR. Furthermore, we have estimated the MCE parameters ($\Delta S_M$ and RC) separately from magnetic and heat capacity measurements and calculated the equivalent adiabatic temperature change ($\Delta T$) indirectly.

## 2. Experimental details

All the samples were prepared in an arc-melting furnace inside a 4N purity argon atmosphere. The ingots were turned and re-melted several times (7-8 times) to ensure homogeneity. 5-8% Mn excess was added during the re-melting process to compensate for the loss during melting. The ingots were then sealed separately in evacuated quartz ampoules and kept at 1173 K for annealing. After 96 hours of heat treatment, the ampoules were quenched in ice water. The room temperature powder X-ray diffraction (XRD) patterns were recorded in Bruker D8 Advance diffractometer using Cu-$K_\alpha$ radiation. Low-temperature XRD was obtained using the powder diffractometer at beamline BL-18B, Photon Factory, KEK, Japan, using an x-ray wavelength of 0.9782 Å. The final compositions were confirmed by energy dispersive X-rays (EDAX) spectrometer (Zeiss Ultra Plus) and given in Table 1 and 2. The temperature and field dependent magnetic properties were measured in a Magnetic Property Measurement System (MPMS, Quantum Design) within the temperature limit 5-320 K and up to 50 kOe fields. The heat capacity was measured in a Physical Property

Measurement System (PPMS, Quantum Design) within the temperature limit 2-300 K. Apiezon N was used as contact grease which get melted above 260 K and therefore, large number of data points were taken to null out the effect of contact grease. The magneto-transport properties were measured using a home-made resistivity/magnetoresistance insert along with an 80 kOe-superconducting magnet system from Oxford Instruments within the temperatures between 5 K and 324 K.

## 3. Results and discussion

*3.1. Structural and temperature dependent magnetic characterization*

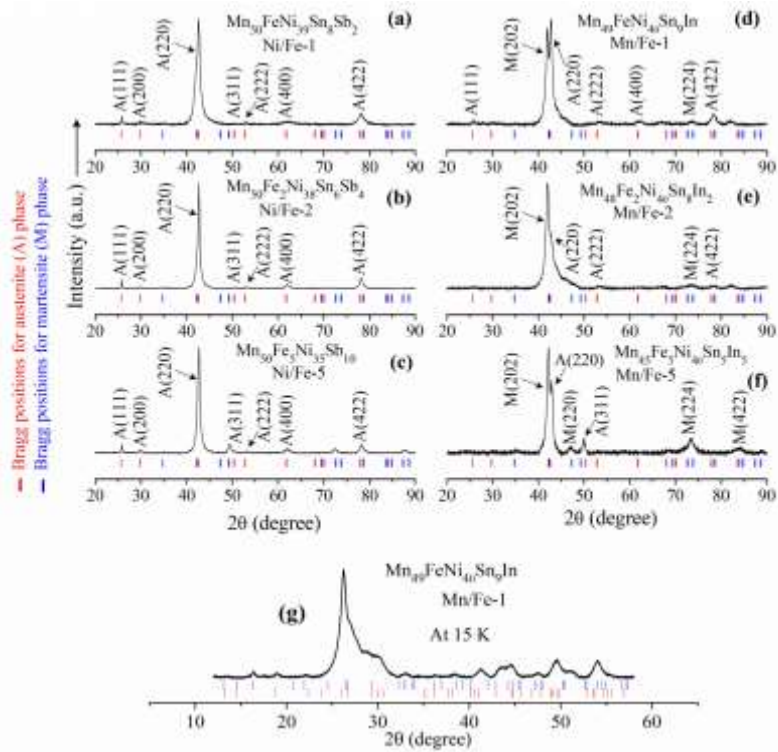

**Fig. 1.** (a-f) Room temperature X-ray (Cu-$K_\alpha$ ; $\lambda$ = 1.5406 Å) diffraction patterns for Mn-Fe-Ni-Sn(In/Sb) alloys. (g) XRD pattern for Mn/Fe-1 at 15 K ($\lambda$ = 0.9782 Å).

Fig. 1(a)-1(f) represents the room temperature XRD patterns for all the samples. The compositions and names of the samples are given in table 1 in detail. The samples of the Ni/Fe series are cubic ($F\bar{4}3m$) at 300 K. The superlattice diffraction peak (111) indicates the increase in atomic order in this series and also predicts that the martensitic transition of these alloys may reside well below room temperature. In the case of Mn/Fe series, Mn/Fe-1 and 2 are in mixed cubic(C)-tetragonal(T) phase whereas the other sample (Mn/Fe-5) is mostly in the tetragonal (*I4/mmm*) phase with volume fraction ratios; C:T ~ 17:3, 7:3 and 1:9 respectively [34]. Fig. 1(g) depicts the XRD pattern for Mn/Fe-1 measured at 15 K. Here, the sample has dominated martensite structure with small mixture of austenite phase due to the generated strain in powder sample [35]. The lattice parameters at 320 K and 15 K have been estimated using Le Bail fit of the temperature dependent XRD data and given in table 1. Ni/Fe-1, 2, Mn/Fe-1 and 2 samples are found to have mostly tetragonal structure at 15 K. As Fe has larger atomic radius than Ni, the cell volume increases slowly with Fe at%. Although, the Fe atoms are smaller than Mn, atomic radius of In is larger than Sn and therefore the lattice parameters increases with increasing Fe content in Mn/Fe series.

**Table 1.** Details of the samples; composition, name, lattice parameters and unit cell volume at 320 K and 15 K.

| Final Composition (Error ~ ±0.1) | Sample's name | 320 K | | | | 15 K | | | |
|---|---|---|---|---|---|---|---|---|---|
| | | a (Å) | b (Å) | c (Å) | volume (Å³) | a (Å) | b (Å) | c (Å) | volume (Å³) |
| $Mn_{50}FeNi_{39}Sn_8Sb_2$ | Ni/Fe-1 | 5.995 | 5.995 | 5.995 | 215.460 | 5.401 | 5.401 | 6.885 | 200.841 |
| $Mn_{50}Fe_2Ni_{38}Sn_6Sb_4$ | Ni/Fe-2 | 5.999 | 5.999 | 5.999 | 215.892 | 5.404 | 5.404 | 6.887 | 201.123 |
| $Mn_{50}Fe_5Ni_{35}Sb_{10}$ | Ni/Fe-5 | 6.021 | 6.021 | 6.021 | 218.276 | ----- | ----- | ----- | ----- |
| $Mn_{49}FeNi_{40}Sn_9In$ | Mn/Fe-1 | 6.050 | 6.050 | 6.050 | 221.445 | 5.454 | 5.454 | 6.903 | 205.337 |
| $Mn_{48}Fe_2Ni_{40}Sn_8In_2$ | Mn/Fe-2 | 6.061 | 6.061 | 6.061 | 222.655 | 5.426 | 5.426 | 6.993 | 205.884 |
| $Mn_{45}Fe_5Ni_{40}Sn_5In_5$ | Mn/Fe-5 | 5.461 | 5.461 | 6.994 | 208.579 | ----- | ----- | ----- | ----- |

The zero field cooled (ZFC) and field cooled cooling (FCC) temperature dependence of magnetizations (*M-T* curves) of all the samples are plotted in Fig. 2 in the presence of 100 Oe magnetic field. The negative slope in the *M-T* curves near 300 K (except for Mn/Fe-5) confirms the existence of ferro-para transition at the Curie temperature of austenite phase ($T_C^A$). Furthermore, the transition with thermal hysteresis between ZFC and FCC curves is a signature of the FOMST in these alloys. The samples of the Mn/Fe series exhibit another phase transition around 140 K which is the Curie temperature of martensite phase ($T_C^M$). The bifurcation of ZFC and FCC *M-T* curves below 125 K indicates the possible existence of mixed ferro-antiferro inter-site interfaces and consequently, an EB effect in the Mn/Fe samples. The characteristic transition temperatures like, austenite start ($A_S$), finish ($A_f$); martensite start ($M_S$) and finish ($M_f$) are obtained by considering the peaks of the 2$^{nd}$ order derivatives of the ZFC, FCC $M$-$T$ curves. Whereas, $T_C^M$ and $T_C^A$ are obtained by considering the peaks of the 1$^{st}$ order derivatives of $M$-$T$ curves. All these temperatures are given in table 2. It is observed that the structural transition temperature decreases when Ni is replaced by Fe, whereas the same increases when Mn is replaced by Fe. It has been reported that the structural transition of these alloys holds a proportional relationship with the $e/a$ ratio [1,3,28–30,36,37]. Here, we have kept the $e/a$ ratio constant by replacing Sn with Sb/In (Table 2). But, the FOMST still shifts as we vary the composition of Mn or Ni by adding Fe. Previously, we have obtained a very similar change in the structural transition temperature in Ni/Fe and Mn/Fe series by varying the total $e/a$ ratio [22]. In the present study, the decrease or increase in FOMST can be attributed to contribution from the valence electron of only the magnetic elements (Ni, Mn, Fe). Therefore, the valence electrons of the post transition elements (Sn, In, Sb) appear to be of less consequence in stabilizing the martensitic transition in this alloy family. $T_C^A$ increases with the increase in Fe and irrespective of $e/a$ ratio for both the series which might due to the enhancement in ferromagnetic correlation with increasing Fe.

**Table 2.** Details of the samples; e/a ratio, austenite start ($A_S$), finish ($A_f$); martensite start ($M_S$), finish ($M_f$), Curie temperatures $T_C^M$ and $T_C^A$.

| Final Composition (Error ~ ±0.1) | Sample's name | $e/a$ total | $e/a$ mag. elements | $A_S$ (K) | $A_f$ (K) | $M_S$ (K) | $M_f$ (K) | $T_C^M$ (K) | $T_C^A$ (K) |
|---|---|---|---|---|---|---|---|---|---|
| $Mn_{50}FeNi_{39}Sn_8Sb_2$ | Ni/Fe-1 | 7.9 | 8.31 | 75 | 146 | 130 | 61 | --- | 288 |
| $Mn_{50}Fe_2Ni_{38}Sn_6Sb_4$ | Ni/Fe-2 | 7.9 | 8.29 | 85 | 140 | 128 | 72 | --- | 290 |
| $Mn_{50}Fe_5Ni_{35}Sb_{10}$ | Ni/Fe-5 | 7.9 | 8.22 | -- | --- | --- | -- | --- | 305 |
| $Mn_{49}FeNi_{40}Sn_9In$ | Mn/Fe-1 | 7.9 | 8.34 | 270 | 288 | 285 | 267 | 135 | 288 |
| $Mn_{48}Fe_2Ni_{40}Sn_8In_2$ | Mn/Fe-2 | 7.9 | 8.36 | 288 | 306 | 300 | 281 | 135 | 294 |
| $Mn_{45}Fe_5Ni_{40}Sn_5In_5$ | Mn/Fe-5 | 7.9 | 8.39 | >320 | >320 | >320 | >320 | 140 | --- |

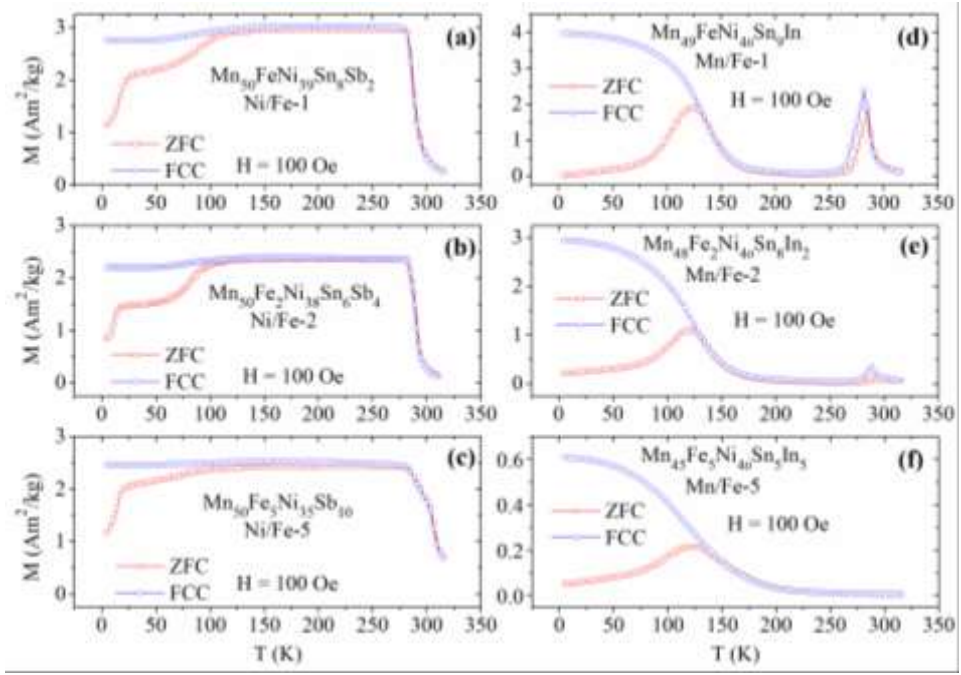

**Fig. 2.** Zero field cooled (ZFC) and field cooled cooling (FC) temperature dependence of magnetization (*M-T* curves) for Mn-Fe-Ni-Sn(In/Sb) alloys in presence of 100 Oe field.

*3.2. Exchange bias effect*

Figs. 3(a)-3(c) represent the FC hysteresis loops for the Mn/Fe series samples within the temperatures between 5 K and 100 K (As the Ni/Fe series does not show EB effect, FC hysteresis loops for them are not shown here). The samples were first cooled from 320 K to the targeted temperature in presence of 10 kOe field and then a full hysteresis loop was measured up to 10 kOe field. One can observe a horizontal shift in the hysteresis loops which confirms the existence of EB behavior in these alloys. A maximum EB field ($H_{EB}$) ~ 775 Oe is obtained for the samples doped with 1 and 2 at% of Fe in the place of Mn (Mn/Fe-1 and Mn/Fe-2). Addition of extra amount of Fe significantly suppresses the exchange anisotropy and $H_{EB}$. Fig. 3(d) shows the temperature dependence of $H_{EB}$ and coercivity ($H_C$) of the Mn/Fe series samples. As the temperature increases, the unidirectional anisotropy created at the interfaces between the ferro and aniferro sites becomes weak due to the decrease in size of the antiferromagnetic matrix and thus $H_{EB}$ decreases with increasing the temperature and almost vanishes near 80 K for all the samples [11–13]. This temperature is known as the EB

blocking temperature ($T_{EB}$). The $H_C$ initially increases with the temperature due to the relative increment of ferromagnetic part in the sample. The same decreases monotonically as the EB effect get blocked.

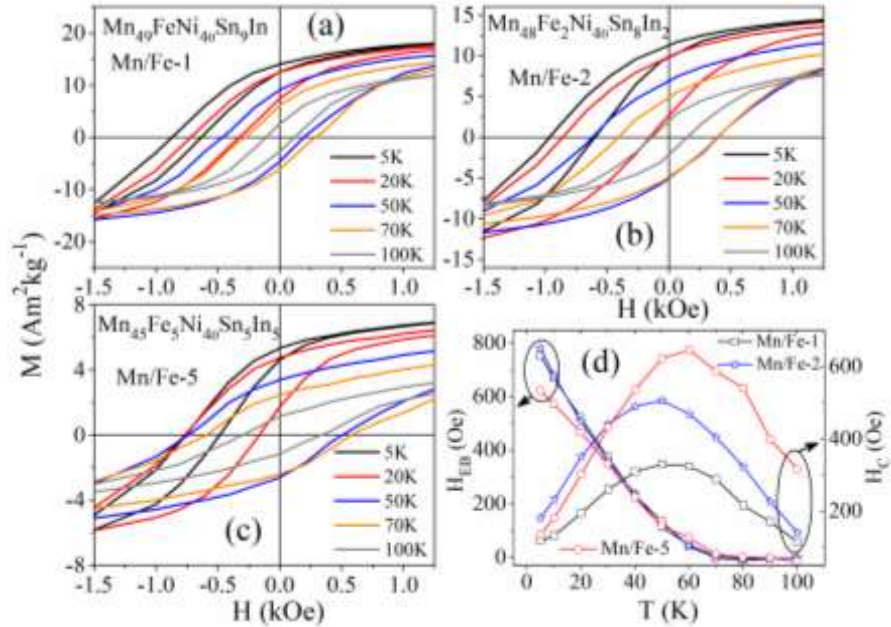

**Fig. 3.** 10 kOe field cooled magnetic hysteresis loops for (a) Mn/Fe-1, (b) Mn/Fe-2 and (c) Mn/Fe-5 samples. (d) Temperature dependent exchange bias ($H_{EB}$) field and coercivity ($H_C$) for Mn/Fe series.

*3.3. MCE from high field M vs T measurement*

The *M-T* curves in presence of 10 kOe and 50 kOe magnetic fields are plotted in Figs. 4(a) – 4(d) for Ni/Fe-1, Ni/Fe-2, Mn/Fe-1 and Mn/Fe-2 samples respectively. A finite shift in the structural transition temperature is observed, which is due to the field induced metamagnetic transition (FIMMT) in these alloys. The structural phases of these samples have a large dependence on the applied magnetic field. It is possible to complete a full transition of structure from tetragonal martensite to cubic austenite phase by only applying a very high magnetic field while keeping the sample at a temperature just below the $A_S$ [38]. The value of saturation magnetization is almost same for Ni/Fe-1 and Ni/Fe-2 (Figs. 4(a) and 4(b)), whereas the same decreases largely when Mn are replaced by Fe (Figs. 4(c), 4(d), Figs. 5(a) and 5(b)). In the former case, replacement of Ni by Fe does not involve significant change in magnetic correlations as the contribution of Ni and Fe in the net magnetic moment of these alloys is very small. On the other hand, decrease in Mn content in the later series causes significant reduction in net magnetization as the Mn atoms carry most of the moment in this alloy family. In addition to that, the structural transition temperature of Mn/Fe-2 almost coincides with the magnetic transition which causes additional suppression in magnetization. Large change in magnetization across the structural transition suggests that a large MCE is likely in such materials. As the structural transition temperature for Mn/Fe-1 and Mn/Fe-2 reside near room temperature, we have calculated the $\Delta S_M$ in them from these high field *M-T* curves by considering the Maxwell's equation [3],

$$\Delta S_M (T, \Delta H) = \mu_0 \sum_H \left( \frac{\partial M}{\partial T} \right)_H \times \Delta H \qquad (1)$$

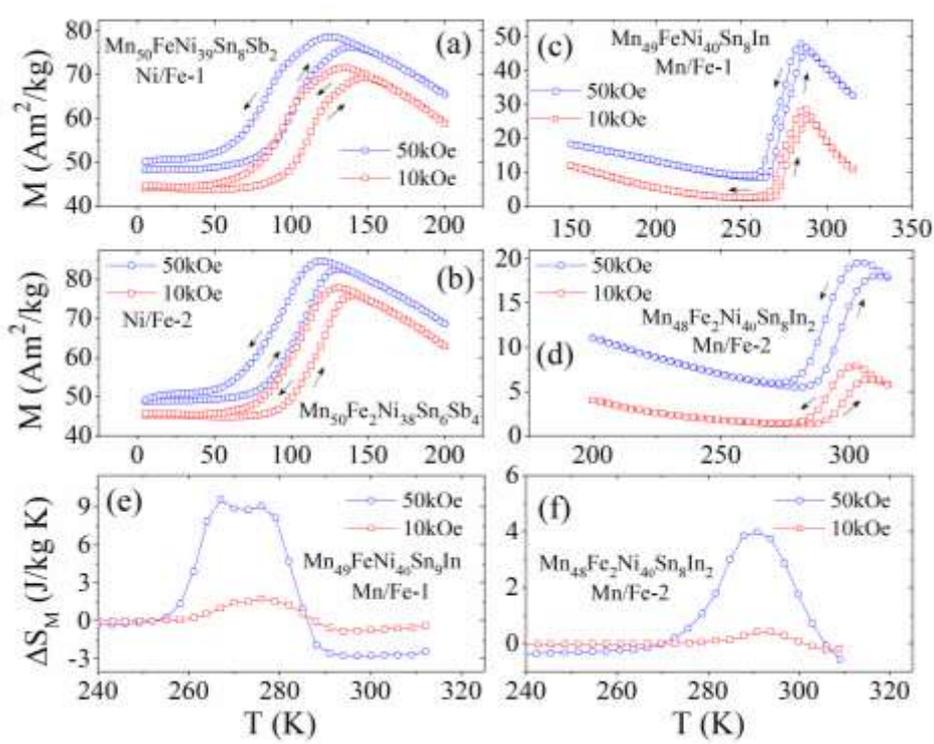

**Fig. 4.** Temperature dependence of magnetization in presence of 10 and 50 kOe fields for (a) Ni/Fe-1, (b) Ni/Fe-2, (c) Mn/Fe-1 and (d) Mn/Fe-2. Temperature dependent magnetic entropy change as calculated from high field *M-T* curves for (e) Mn/Fe-1 and (f) Mn/Fe-2 during cooling.

where, $\mu_o$, $M$, $T$ and $H$ are the permeability of free space, sample's magnetization, measurement temperature and applied magnetic field respectively. Figs. 4(e) and 4(f) represent the temperature dependent plot of $\Delta S_M$ as calculated from high field $M$-$T$ curves during cooling ($M$-$T$ curves under 20, 30 and 40 kOe were also measured and used to calculate the $\Delta S_M$ from equation (1), but not presented in Fig. 4 for the sake of brevity). One can find a large MCE ($\Delta S_M \sim 10$ Jkg$^{-1}$K$^{-1}$) with a table like temperature dependency for Mn/Fe-1 sample.

*3.4. MCE from isothermal M vs H measurement*

Figs. 5(a) and 5(b) represent the isothermal field dependence of magnetization ($M$-$H$ curves) for Mn/Fe-1 and Mn/Fe-2 samples across their respective FOMST. We have chosen these two samples for MCE investigations as their structural transition temperature is near 300 K, and thus could be of the interest for room temperature magnetic refrigeration. The $M$-$H$ curves are measured during cooling in the temperatures between $M_S$ and $M_f$. The FOMST of these alloys significantly suffers from the field history effect [38–40]. Therefore, we have followed the discontinuous cooling protocol where the sample was heated to 320 K and cooled back to the targeted temperature before taking data. The magnetization at 50 kOe field decreases with the measurement temperature due to the structural phase transition. The $\Delta S_M$ of these two alloys has also been calculated from the isothermal $M$-$H$ curves using the Maxwell's equation [3]

$$\Delta S_M(T, \Delta H) = \mu_0 \int_0^{H_{max}} \left( \frac{\partial M}{\partial T} \right)_H dH \tag{2}$$

Figs. 5(c) and 5(d) show the temperature dependent plot of $\Delta S_M$ for Mn/Fe-1 and Mn/Fe-2 samples up to a field change of 50 kOe. A very similar table like behavior [41] is observed in the $\Delta S_M$–$T$ curves of Mn/Fe-1 as we have already observed in Fig. 4(e) ) in Sec. 3.3. Such nature may occur if there is a strong field induced effect in the sample, where the structure gets linearly dragged to the magnetically more ordered phase upon applying magnetic field. If we consider the practical applicability of these materials, such table like curve is good for achieving large cooling power with an extended working temperature region. The peak value of $\Delta S_M$ decreases for Mn/Fe-2 due to two main reasons; *i)* total magnetization of the austenite phase decreases with decreasing Mn and *ii)* the $T_C^A$ and structural transition almost coincide, which also suppresses the net magnetization and thus the d*M/*d*T*.

**Table 3.** Magnetocaloric parameters for Mn/Fe series samples estimated under different measurements.

| Composition | Sample's name | $\Delta S_M^{Peak}$ (Jkg$^{-1}$K$^{-1}$) for $\Delta H$ = 50kOe | | | RC (Jkg$^{-1}$) for $\Delta H$ = 50kOe | | |
|---|---|---|---|---|---|---|---|
| | | *Isofield M* vs *T* | *Isothermal M* vs *H* | *Heat Capacity* | *Isofield M* vs *T* | *Isothermal M* vs *H* | *Heat Capacity* |
| $Mn_{49}FeNi_{40}Sn_9In$ | Mn/Fe-1 | 9.57 | 9.45 | 8.1 | 190.73 | 147.35 | 140.2 |
| $Mn_{48}Fe_2Ni_{40}Sn_8In_2$ | Mn/Fe-2 | 3.97 | 3.38 | --- | 66.93 | 45.05 | --- |

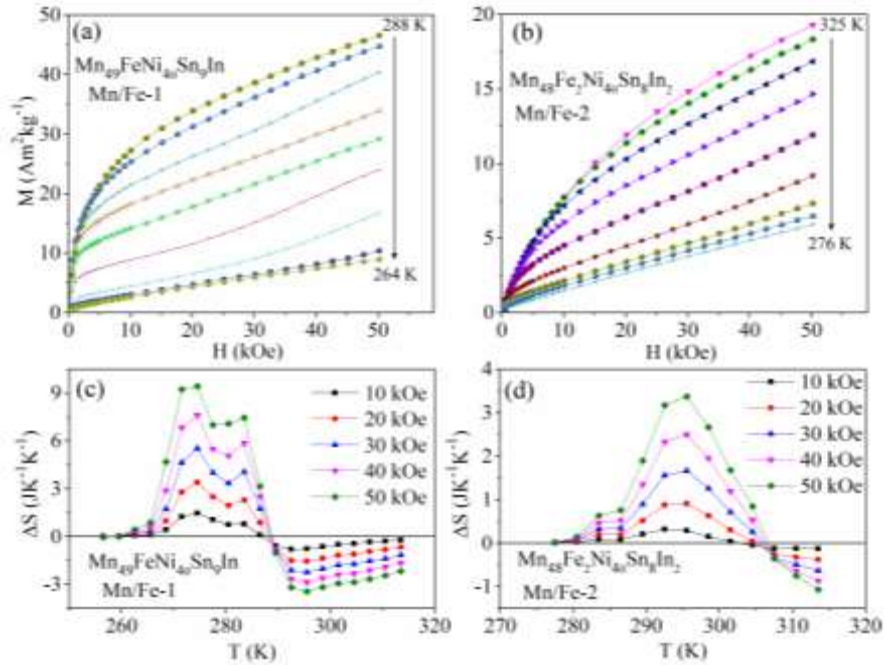

**Fig. 5.** Isothermal field dependence of magnetization (*M-H* curves) for (a) Mn/Fe-1 and (b) Mn/Fe-2. Temperature dependence of magnetic entropy change ($\Delta S_M$ –*T* curves) as calculated from the *M-H* curves for (c) Mn/Fe-1 and (d) Mn/Fe-2.

*3.5. MCE from heat capacity measurements*

We have also verified the MCE of Mn/Fe-1 by heat capacity (*HC*) measurements. Fig. 6 shows the temperature dependence of *HC* in the presence of zero and 50 kOe fields where the exothermic peak confirms the existence of FOMST. The inset shows the $\Delta S_M$ –*T* curves as measured from the *HC* data by using the formula [2,3]

$$\Delta S_M(T, \Delta H) = \int_{T_0}^{T} \frac{(HC_{50kOe} - HC_0)}{T} dT \tag{3}$$

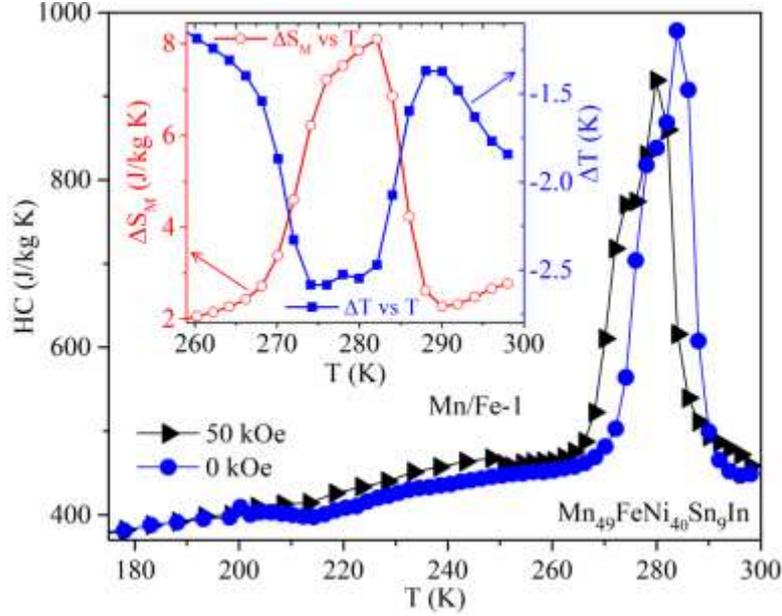

**Fig. 6.** Temperature dependence of heat capacity in presence of zero and 50 kOe fields for Mn/Fe-1 sample. **Inset:** Temperature dependence of magnetic entropy change and equivalent indirect temperature change as calculated from the heat capacity data for Mn/Fe-1.

where, $T_0$ denotes the lowest obtainable temperature of our measurement ( ~ 2 K). As we are only calculating the change in entropy, this assumption of integration limit will not affect the result. The peak value of $\Delta S_M$ is lower compared to that obtained from magnetization data (table 3). The possible reason is the presence of the FIMMT which causes unwanted spikes in the $\Delta S_M$ –T curve as plotted from magnetization data (Figs. 4(e) and 5(d) in Sec. 3.4). However, the peak is similarly broadened for Mn/Fe-1 sample. Therefore, we can achieve a large cooling power from all the $\Delta S_M$ –T curves of the same sample. One can notice that the values of magnetocaloric parameters obtained from the three different measurement methods as reported in this paper are not considerably different with one another. In more general words, magnetic refrigeration technology holds universality in terms of different investigation methodologies. We have also used the *HC* data to calculate the equivalent indirect change in temperature ($\Delta T$) by using equation (4) where, $T_H$ and $T_0$ are respectively the temperatures for $H = 50$ kOe and 0 under constant entropy or adiabatic condition. The maximum $\Delta T$ is found to be -2.6 K for the Mn/Fe-1 sample near 274 K. Such negative value of $\Delta T$ corresponds to the adiabatic magnetization process where magnetic refrigeration can be realized by adiabatically applying the magnetic field.

$$\Delta T(T, \Delta H) = (T_H - T_0)_S \qquad (4)$$

We have calculated the refrigerant capacity (RC) using equation (5) [40] from all the $\Delta S_M - T$ curves as estimated from isofield $M$ vs $T$, isothermal $M$ vs $H$ and HC measurements. The estimated values are tabulated in table 3. One can find that a large amount of RC is available due to broadened table like peak in the $\Delta S_M - T$ curves. The RC values as calculated from isothermal $M$ vs $H$ measurements are very close to that obtained from the HC measurements.

$$RC = \int_{T_1}^{T_2} |\Delta S_M(T)|\, dT \qquad (5)$$

### 3.6. Magnetoresistance

The magnetoresistance of these materials has been estimated using equation (6) from zero field and 80 kOe temperature dependent resistivity data during heating and cooling which is plotted in Fig. $\rho_0(T)$ and $\rho_H(T)$ correspond to respectively the zero field and in-field resistivity values at temperature $T$. Nearly, 30% MR is obtained for Mn/Fe-1 sample. A large drop in resistivity across the structural transition and its field induced shift result in a large negative MR around transition temperature. In the case of Ni/Fe-2 more than 20 % MR is observed over a broad temperature range below 75 K. It could be attributed to incomplete martensitic transition during cooling in the presence of 80 kOe magnetic field. Similar MR behavior has been observed in $Ni_2Mn_{1.36}In_{0.64}$ by Singh et al. [42], where they showed large difference in zero field cooled and field cooled MR. Earlier magnetization studies have shown that austenite to martensite transition is arrested at low temperature and high magnetic field e.g. Sharma et al. in $Ni_{50}Mn_{34}In_{16}$ [43] and Banerjee et al. in $Ni_{45}Co_5Mn_{38}Sn_{12}$ [44]. The presence of kinetic arrest results in path dependent magnetic state and hence magnetoresistance.

$$MR(T,\Delta H) = \frac{(\rho_H(T) - \rho_0(T))}{\rho_o(T)} \times 100 \tag{6}$$

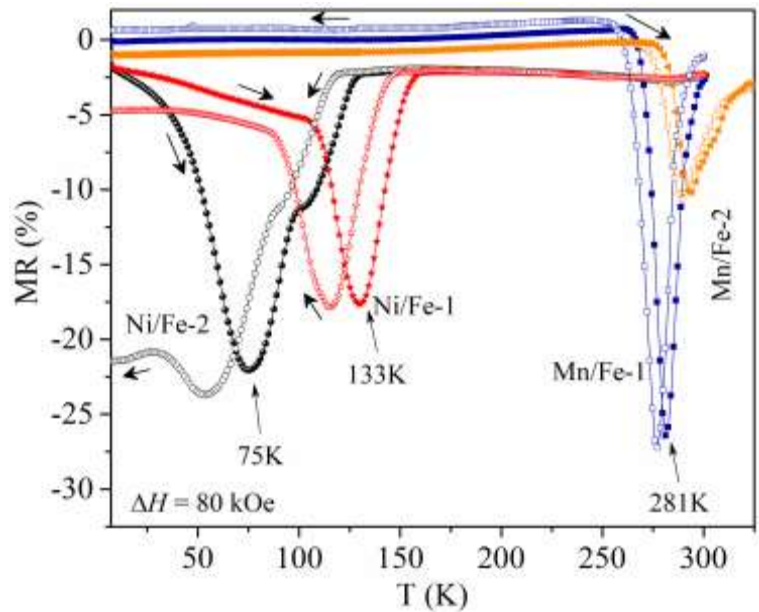

**Fig. 7.** Temperature dependence of magnetoresistance under a field change of 80 kOe. Open and closed symbols represent the cooling and heating data respectively.

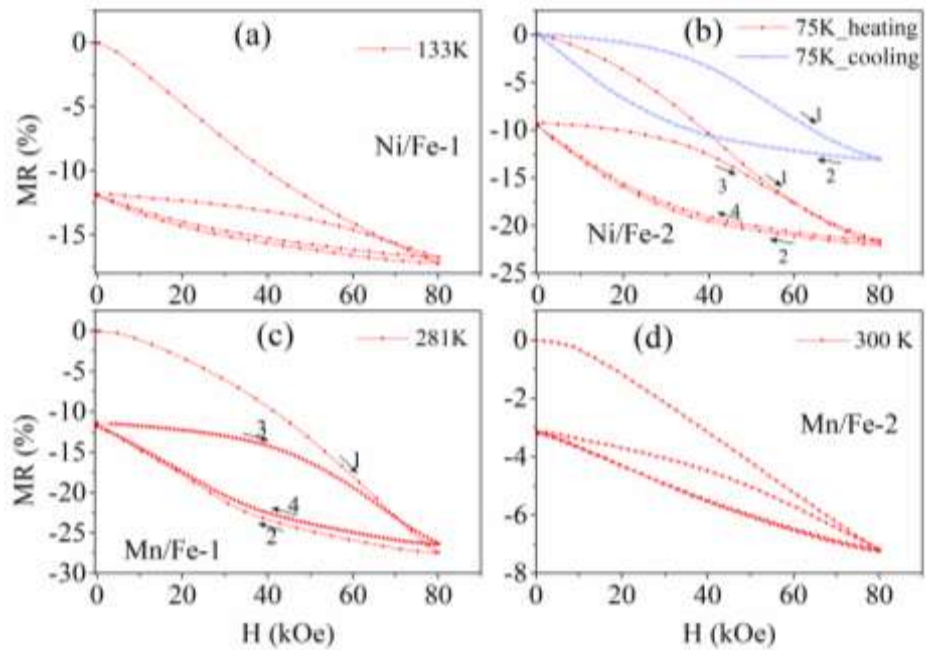

**Fig. 8.** Isothermal field dependence of magnetoresistance for (a) Ni/Fe-1, (b) Ni/Fe-2, (c) Mn/Fe-1 and (d) Mn/Fe-2 at their respective structural transition temperatures.

Isothermal MR is plotted in Fig. 8 for Ni/Fe-1, Ni/Fe-2, Mn/Fe-1 and Mn/Fe-2 samples. For the curves shown in red color, ZFC state is obtained by cooling the sample to 5K and then warmed up to the measurement temperature under zero field. These measurements show large negative MR due to field induced martensite to austenite transformation (curve-1). For curve-2 (80 to 0 kOe) MR do not returns back to zero resulting in an open loop i.e. only a part of field induced austenite phase transforms back to martensite. The measurement temperatures in Fig. 8 are within the thermal hysteresis region of zero field resistivity of the respective samples. Therefore, the origin of open loop could be due to metastable supercooled and superheated state. For such metastable states, presence of open loop in MR depends on the history of sample before the application of magnetic field [45]. For the present sample critical field vs temperature curve has negative slope (or transition temperature decreases with magnetic field) and therefore open loop in MR is expected for measurement temperature reached by warming but not during cooling. During warming a fraction of martensite phase exist as superheated metastable state and it transform to stable austenite phase with the isothermal application of magnetic field. On reducing magnetic field to zero, system remains in austenite state due to the presence of free energy barrier. On the other hand for ZFC state from 300 K, there will be no metastable martensite state and hence MR will reach zero after field cycling. A representative measurement to verify it is shown in Fig. 8(b) in blue color where 75 K is reached by cooling from 300 K. In case of isothermal magnetization measurement similar history dependence results in virgin curve lying outside the envelope curve, which has been demonstrated in $Ni_{50}Mn_{34}In_{16}$ by Sharma et al. [43].

## 4. Conclusion

In summary, we have performed a systematic study on the magneto-structural transition, exchange bias, magnetocaloric effect (MCE) and magnetoresistance in Mn-rich Fe-doped Mn-Fe-Ni-Sn(Sb/In) Heusler alloys. The martensitic transition temperature has been found to shift (increase or decrease) proportionally with the valence electron concentration of only the magnetic elements present in the samples. The estimation of MCE from high field temperature dependent magnetic measurement (*M-T* curves) may overestimate the results in terms of cooling power. However we observe that the calculations of MCE from isothermal field dependent magnetic data (*M-H* curves) and heat capacity data match well with each other. The broadened working region and a large cooling power ~ 140 Jkg$^{-1}$ with -2.6 K indirect temperature change can make the Mn$_{49}$FeNi$_{40}$Sn$_9$In alloy a potential refrigerant for room temperature magnetic refrigeration.

**Acknowledgements**

Arup Ghosh is thankful to SERB, DST, Govt. of India for providing the financial support through National Post-Doctoral Fellowship (PDF/2015/000599).  SN acknowledges funding support by the Department of Science and Technology (DST, Govt. of India) under the DST Nanomission Thematic Unit Program (SR/NM/TP-13/2016). We thank Vikram Singh, Pampi Saha from UGC-DAE Consortium for Scientific Research, Indore and Devendra D Buddhikot from Tata Institute of Fundamental Research, Mumbai for their help during magneto-transport and heat capacity measurements respectively.

**References**

[1]  T. Krenke, E. Duman, M. Acet, E.F. Wassermann, X. Moya, L. Manosa, A. Planes, Inverse magnetocaloric effect in ferromagnetic Ni-Mn-Sn alloys, Nat Mater. 4 (2005) 450–454. http://dx.doi.org/10.1038/nmat1395.

[2]  A. Planes, L. Mañosa, M. Acet, Magnetocaloric effect and its relation to shape-memory properties in ferromagnetic Heusler alloys., J. Phys. Condens. Matter. 21 (2009) 1–29. doi:10.1088/0953-8984/21/23/233201.

[3]  V.D. Buchelnikov, V. V Sokolovskiy, Magnetocaloric Effect in Ni–Mn–X(X=Ga, In, Sn, Sb) Heusler Alloys, Phys. Met. Metallogr. 112 (2011) 633–665. doi:10.1134/S0031918X11070052.

[4]  M. Kohl, V. Chernenko, D. Bourgault, J. Tillier, P. Courtois, X. Chaud, N. Caillault, L. Carbone, 3rd International Symposium on Shape Memory Materials for Smart Systems/E-MRS 2010 Spring Meeting Large Magneto-Caloric Effect in Ni–Co–Mn–In systems at room temperature, Phys. Procedia. 10 (2010) 120–124. doi:http://dx.doi.org/10.1016/j.phpro.2010.11.086.

[5]  J.-H. Chen, N.M. Bruno, I. Karaman, Y. Huang, J. Li, J.H. Ross, Calorimetric and magnetic study for Ni50Mn36In14 and relative cooling power in paramagnetic inverse magnetocaloric systems, J. Appl. Phys. 116 (2014). doi:http://dx.doi.org/10.1063/1.4902527.

[6]  W. Ito, K. Ito, R.Y. Umetsu, R. Kainuma, K. Koyama, K. Watanabe, A. Fujita, K. Oikawa, K. Ishida, T. Kanomata, Kinetic arrest of martensitic transformation in the NiCoMnIn metamagnetic shape memory alloy, Appl. Phys. Lett. 92 (2008) 1–4. doi:10.1063/1.2833699.

[7]  A. Ghosh, K. Mandal, Large magnetoresistance associated with large inverse magnetocaloric effect in Ni-Co-Mn-Sn alloys, Eur. Phys. J. B. 579 (2013) 295–299. doi:10.1140/epjb/e2013-40489-0.

[8]  B. Zhang, X.X. Zhang, S.Y. Yu, J.L. Chen, Z.X. Cao, G.H. Wu, Giant magnetothermal conductivity in the Ni–Mn–In ferromagnetic shape memory alloys, Appl. Phys. Lett. 91 (2007) 12510. doi:10.1063/1.2753710.

[9]  T. Sakon, N. Fujimoto, T. Kanomata, Y. Adachi, Magnetostriction of Ni2Mn1−xCrxGa Heusler Alloys, Metals (Basel). 7 (2017) 410. doi:10.3390/met7100410.

[10] B.S. Choi, Thermal, magnetic, and magnetoelastic data on three different Heusler

alloys based on Ni-Mn-X (X=Ga, in, or Sn), IEEE Trans. Magn. 42 (2006) 1770–1777. doi:10.1109/TMAG.2006.874305.

[11] M. Khan, I. Dubenko, S. Stadler, N. Ali, Exchange bias behavior in Ni–Mn–Sb Heusler alloys, Appl. Phys. Lett. 91 (2007) 072510. doi:http://dx.doi.org/10.1063/1.2772233.

[12] Z. Li, C. Jing, J. Chen, S. Yuan, S. Cao, J. Zhang, Observation of exchange bias in the martensitic state of Ni50Mn36Sn14 Heusler alloy, Appl. Phys. Lett. 91 (2007) 112505. doi:http://dx.doi.org/10.1063/1.2784958.

[13] M. Khan, I. Dubenko, S. Stadler, N. Ali, Exchange bias in bulk Mn rich Ni–Mn–Sn Heusler alloys, J. Appl. Phys. 91 (2007) 113914. doi:http://dx.doi.org/10.1063/1.2818016.

[14] B. Wang, Y. Liu, Exchange Bias and Inverse Magnetocaloric Effect in Co and Mn Co-Doped Ni2MnGa Shape Memory Alloy, Metals (Basel). 3 (2013) 69–76. doi:10.3390/met3010069.

[15] T. Krenke, E. Duman, M. Acet, X. Moya, L. Mañosa, A. Planes, Effect of Co and Fe on the inverse magnetocaloric properties of Ni-Mn-Sn, J. Appl. Phys. 102 (2007) 033903. doi:http://dx.doi.org/10.1063/1.2761853.

[16] A.K. Nayak, K.G. Suresh, A.K. Nigam, Giant inverse magnetocaloric effect near room temperature in Co substituted NiMnSb Heusler alloys, J. Phys. D. Appl. Phys. 42 (2009) 35009. http://stacks.iop.org/0022-3727/42/i=3/a=035009.

[17] A.K. Nayak, K.G. Suresh, A.K. Nigam, Irreversibility of field-induced magnetostructural transition in NiCoMnSb shape memory alloy revealed by magnetization, transport and heat capacity studies, Appl. Phys. Lett. 96 (2010) 2010–2013. doi:10.1063/1.3365181.

[18] R.Y. Umetsu, A. Sheikh, W. Ito, B. Ouladdiaf, K.R.A. Ziebeck, T. Kanomata, R. Kainuma, The effect of Co substitution on the magnetic properties of the Heusler alloy Ni50Mn33Sn17, Appl. Phys. Lett. 98 (2011) 10–13. doi:10.1063/1.3548558.

[19] Z. Han, J. Chen, B. Qian, P. Zhang, X. Jiang, D. Wang, Y. Du, Phase diagram and magnetocaloric effect in Mn2Ni1.64−xCoxSn0.36 alloys, Scr. Mater. 66 (2012) 121–124. doi:10.1016/j.scriptamat.2011.10.020.

[20] A. Ghosh, K. Mandal, Large inverse magnetocaloric effect in Ni48.5−xCoxMn37Sn14.5 (x=0, 1 and 2) with negligible hysteresis, J. Alloys Compd. 579 (2013) 295–299. doi:10.1016/j.jallcom.2013.06.062.

[21] F. Chen, Y.X. Tong, B. Tian, L. Li, Y.F. Zheng, Y. Liu, Magnetic-field-induced

reverse transformation in a NiCoMnSn high temperature ferromagnetic shape memory alloy, J. Magn. Magn. Mater. 347 (2013) 72–74. doi:http://dx.doi.org/10.1016/j.jmmm.2013.07.049.

[22] A. Ghosh, K. Mandal, Effect of Fe substitution on the magnetic and magnetocaloric properties of Mn-rich Mn-Ni-Fe-Sn off-stoichiometric Heusler alloys, J. Appl. Phys. 117 (2015) 093909.

[23] F. Chen, Y.-X. Tong, B. Tian, L. Li, Y.-F. Zheng, Martensitic transformation and magnetic properties of Ti-doped NiCoMnSn shape memory alloy, Rare Met. 33 (2014) 511–515. doi:10.1007/s12598-013-0100-7.

[24] J. Kamarád, F. Albertini, Z. Arnold, S. Fabbrici, J. Kaštil, Extraordinary magnetic and structural properties of the off-stoichiometric and the Co-doped Ni2MnGa Heusler alloys under high pressure, Acta Mater. 77 (2014) 60–67. doi:http://dx.doi.org/10.1016/j.actamat.2014.06.011.

[25] Y. Jiang, Z. Li, Z. Li, Y. Yang, B. Yang, Y. Zhang, C. Esling, X. Zhao, L. Zuo, Magnetostructural transformation and magnetocaloric effect in Mn-Ni-Sn melt-spun ribbons, Eur. Phys. J. Plus. 132 (2017) 42. doi:10.1140/epjp/i2017-11316-1.

[26] Z. Li, Y. Jiang, Z. Li, C.F. Sánchez Valdés, J.L. Sánchez Llamazares, B. Yang, Y. Zhang, C. Esling, X. Zhao, L. Zuo, Phase transition and magnetocaloric properties of Mn50Ni42-xCoxSn8 (0 ≤ x ≤ 10) melt-spun ribbons, IUCrJ. 5 (2018) 54–66. doi:10.1107/S2052252517016220.

[27] L.H. Yang, H. Zhang, F.X. Hu, J.R. Sun, L.Q. Pan, B.G. Shen, Magnetocaloric effect and martensitic transition in Ni50Mn36−xCoxSn14, J. Alloys Compd. 588 (2014) 46–48. doi:https://doi.org/10.1016/j.jallcom.2013.10.196.

[28] L. Ma, S.Q. Wang, Y.Z. Li, C.M. Zhen, D.L. Hou, W.H. Wang, J.L. Chen, G.H. Wu, Martensitic and magnetic transformation in Mn50Ni50−xSnx ferromagnetic shape memory alloys, J. Appl. Phys. 112 (2012) 83902. doi:10.1063/1.4758180.

[29] S.E. Muthu, N.V.R. Rao, M.M. Raja, D.M.R. Kumar, D.M. Radheep, S. Arumugam, Influence of Ni/Mn concentration on the structural, magnetic and magnetocaloric properties in Ni 50− x Mn 37+ x Sn 13 Heusler alloys, J. Phys. D. Appl. Phys. 43 (2010) 425002. doi:10.1088/0022-3727/43/42/425002.

[30] L. Ma, S.Q. Wang, Y.Z. Li, C.M. Zhen, D.L. Hou, W.H. Wang, J.L. Chen, G.H. Wu, Martensitic and magnetic transformation in Mn50Ni50−xSnx ferromagnetic shape memory alloys, J. Appl. Phys. 112 (2012) 083902. doi:10.1063/1.4758180.

[31] H.C. Xuan, Y.X. Zheng, S.C. Ma, Q.Q. Cao, D.H. Wang, Y.W. Du, The martensitic

transformation, magnetocaloric effect, and magnetoresistance in high-Mn content Mn47+xNi43−xSn10 ferromagnetic shape memory alloys, J. Appl. Phys. 108 (2010) 103920. doi:10.1063/1.3511748.

[32] Q. Tao, Z.D. Han, J.J. Wang, B. Qian, P. Zhang, X.F. Jiang, D.H. Wang, Y.W. Du, Phase stability and magnetic-field-induced martensitic transformation in Mn-rich NiMnSn alloys, AIP Adv. 2 (2012) 42181. doi:10.1063/1.4772626.

[33] J.S. Amaral, V.S. Amaral, The effect of magnetic irreversibility on estimating the magnetocaloric effect from magnetization measurements, Appl. Phys. Lett. 94 (2009) 42506. doi:10.1063/1.3075851.

[34] A.K. De, D.C. Murdock, M.C. Mataya, J.G. Speer, D.K. Matlock, Quantitative measurement of deformation-induced martensite in 304 stainless steel by X-ray diffraction, Scr. Mater. 50 (2004) 1445–1449. doi:https://doi.org/10.1016/j.scriptamat.2004.03.011.

[35] F. Alvarado-Hernández, O. Jiménez, G. González-Castañeda, V. Baltazar-Hernández, J. Cabezas-Villa, M. Albiter, H. Vergara-Hernández, L. Olmos, Sintering kinetics of Ni2FeSb powder alloys produced by mechanical milling, Trans. Nonferrous Met. Soc. China. 26 (2016) 2126–2135. doi:https://doi.org/10.1016/S1003-6326(16)64326-1.

[36] H.C. Xuan, Y.X. Zheng, S.C. Ma, Q.Q. Cao, D.H. Wang, Y.W. Du, The martensitic transformation, magnetocaloric effect, and magnetoresistance in high-Mn content Mn47+xNi43−xSn10 ferromagnetic shape memory alloys, J. Appl. Phys. 108 (2010) 103920. doi:10.1063/1.3511748.

[37] Q. Tao, Z.D. Han, J.J. Wang, B. Qian, P. Zhang, X.F. Jiang, D.H. Wang, Y.W. Du, Phase stability and magnetic-field-induced martensitic transformation in Mn-rich NiMnSn alloys, AIP Adv. 2 (2012) 42181. doi:10.1063/1.4772626.

[38] P.J. Shamberger, F.S. Ohuchi, Hysteresis of the martensitic phase transition in magnetocaloric-effect Ni-Mn-Sn alloys, Phys. Rev. B - Condens. Matter Mater. Phys. 79 (2009) 1–9. doi:10.1103/PhysRevB.79.144407.

[39] L. Caron, Z.Q. Ou, T.T. Nguyen, D.T.C. Thanh, O. Tegus, E. Brück, On the determination of the magnetic entropy change in materials with first-order transitions, J. Magn. Magn. Mater. 321 (2009) 3559–3566. doi:https://doi.org/10.1016/j.jmmm.2009.06.086.

[40] A. Ghosh, P. Sen, K. Mandal, Measurement protocol dependent magnetocaloric properties in a Si-doped Mn-rich Mn-Ni-Sn-Si off-stoichiometric Heusler alloy, J. Appl. Phys. 119 (2016) 183902. doi:10.1063/1.4948962.

[41] L.Y. Ma, L.H. Gan, K.C. Chan, D. Ding, L. Xia, Achieving a table-like magnetic entropy change across the ice point of water with tailorable temperature range in Gd-Co-based amorphous hybrids, J. Alloys Compd. 723 (2017) 197–200. doi:https://doi.org/10.1016/j.jallcom.2017.06.254.

[42] S. Singh, I. Glavatskyy, C. Biswas, Field-cooled and zero-field cooled magnetoresistance behavior of Ni2Mn1+xIn1−x alloys, J. Alloys Compd. 615 (2014) 994–997. doi:https://doi.org/10.1016/j.jallcom.2014.07.012.

[43] V.K. Sharma, M.K. Chattopadhyay, S.B. Roy, Kinetic arrest of the first order austenite to martensite phase transition in Ni50Mn34In16: dc magnetization studies, Phys. Rev. B. 76 (2007) 140401. doi:10.1103/PhysRevB.76.140401.

[44] A. Banerjee, P. Chaddah, S. Dash, K. Kumar, A. Lakhani, X. Chen, R. V Ramanujan, History-dependent nucleation and growth of the martensitic phase in the magnetic shape memory alloy Ni45Co5Mn38Sn12, Phys. Rev. B. 84 (2011) 214420. doi:10.1103/PhysRevB.84.214420.

[45] P. Kushwaha, R. Rawat, P. Chaddah, Metastability in the ferrimagnetic–antiferromagnetic phase transition in Co substituted Mn 2 Sb, J. Phys. Condens. Matter. 20 (2008) 22204. http://stacks.iop.org/0953-8984/20/i=2/a=022204.